\documentclass[12pt]{article}
\usepackage{saosutab}

\advance\textheight by\topskip

\begin{document}

\title{\bf Low frequency catalogues of the CATS database}

\author {{\bf O.V.Verkhodanov$^1$,~S.A.Trushkin$^1$,~H.Andernach$^2$}
\and {\small $^1$ Special Astrophysical Observatory, Nizhnij Arkhyz, Russia,
357147}
\and {\small $^2$ Depto.\,de Astronom\'\i a, Apdo.\ Postal 144,
 Univ.\ Guanajuato, Guanajuato, Mexico}
\and {\it E-mail:~vo@sao.ru,~satr@sao.ru,~heinz@astro.ugto.mx}
}
\date{}
\maketitle

\begin{abstract}
The authors describe the largest existing publicly accessible radio
source database CATS (``astrophysical CATalogues Supporting system''). CATS
contains more than 300 catalogues of objects
detected in different (but mostly radio) wavelength ranges. These include
catalogues from the largest existing surveys, like NVSS, FIRST, WENSS,
TXS, GB6, IRAS, ROSAT, PGC, MCG, etc. Thus CATS allows to draw samples of
objects to study a great variety of astrophysical problems. \\
CATS also includes all of the largest low-frequency catalogues published so far\,:
UTR(10--25~MHz), Culgoora (80, 160\,MHz), Cambridge 
(38, 151, 178\,MHz), Miyun (232\,MHz), WENSS (325\,MHz), Texas (365\,MHz), 
Bologna and Molonglo (408\,MHz) etc., and allows the user to operate 
with them.
\end{abstract}

The creation of the CATS database ({\bf http://cats.sao.ru}) 
was motivated on the one hand by the measurements of radio continuum 
spectra with the RATAN-600 telescope (by two of us, O.V. and S.T.), and,
on the other hand, by the activity of the third author (H.A.) in collecting
the astrophysical catalogues.  CATS became operational in 1996.

The main objective of the CATS database is to help a user to operate
with a large number of astrophysical catalogues which have been published in
different formats, each one including a somewhat different set of 
observational parameters.

Among the options accessible to CATS users we mention
\begin{itemize}
\item
    Request short descriptions of each catalogue, or print a list of
    catalogues covering the required sky areas.
\item
    Select objects from one or more catalogues by coordinate,
    flux, spectral index, frequency, etc.
\item
    Select objects from one or more catalogues for many sky patches
    defined by position and size (e.g.\ for cross-ID with other catalogues).
\item
    Display radio spectra of selected sources.
\end{itemize}

Among other applications, CATS may be used for object identification, 
source counts in different wavelength ranges, or the study of statistical 
properties of different source populations.
Standard access to CATS is via WWW at {\tt\small http://cats.sao.ru}), 
or e-mail (cats@sao.ru), or anon.\ FTP.  Batch requests to 
extract objects from large lists of sky regions 
are supported via e-mail.
Standard formats of input and output of CATS are discussed,
and some optimized ways of interaction with CATS are shown.

Our database comprises more than 300 catalogues. At present, 140 of these
catalogues include data at frequencies less or equal 1400~MHz. The largest or
the most important of them are all the Cambridge catalogues from
3C to 8C (38, 151, 178, 408~MHz),
the VLA FIRST and NVSS catalogues at 1400~MHz, the 2$^{nd}$ and 3$^{rd}$ Bologna
surveys at 408~MHz, the Westerbork catalogues at 325~MHz,
the Texas catalogue at 365~MHz,
the Parkes catalogues at several frequencies, the GB catalogues, the RATAN-600
catalogues. There are further dedicated catalogues of radio sources, like e.g.\
those in clusters of galaxies, by Slee, Owen or Reynolds. The CATS database 
includes a large number of Galactic radio sources, supernova remnants and 
their maps, observed with the RATAN-600, Bonn, Ooty, VLA and other telescopes.

Several basic catalogues at low frequencies are given in Table, containing
the main survey parameters. This table also includes catalogus obtained
at high radio frequencies, at other wavelength ranges and several mixed
catalogues.


The standard access to the CATS database is provided in three different ways:
\begin{enumerate}
\item
    http://cats.sao.ru
\item
    e-mail:~cats@sao.ru  (an empty email delivers a help file)
\item
    ftp://cats.sao.ru  (anonymous FTP)
\end{enumerate}

HTTP and e-mail access permits the user to obtain informtaion from catalogues
in selected areas, and to search for cross-identifications of entries in CATS 
with user-specified input lists.

Here are some simple examples of e-mail requests which are very convenient
for a slow network transfer:

\begin{enumerate}
\item
    Search an area defined by limits in equatorial coordinates for objects
drawn from all radio catalogues, with a flux density higher than 500~mJy:

{\tt
mail cats@sao.ru\\
cats select\\
 ra min=12:30 max=12:40:15.  dec $>$ 0 $<$ 5' 46"\\
 catalogs r epoch=1950  flux $>$ 500\\
cats end
}
\item  Request letter for the coordinate cross-identification
 of 3 sources in all the CATS catalogs with errors by x (RA) and y (DEC):\\
{\tt
mail cats@sao.ru\\
cats match catalogs a\\
 window box x=60" y=2'\\
 sources:\\
 s1 02:02:00~~~ +31:23:16~~~ 1950\\
 s2 02:23:10~~~~~00:03:00~~~~1950\\
 s3 21:26:33.9~~-18:34:33.0~ 1950\\
cats end
}
\end{enumerate}

{
\large
\begin{center}
\topcaption {\bf
Basic catalogues of the CATS data base}

\tablehead{\hline
	\multicolumn{1}{|c|}{Freq}
      & \multicolumn{1}{|c|}{ }
      & \multicolumn{1}{|c|}{Year}
      & \multicolumn{1}{|c|}{RA(h)}
      & \multicolumn{1}{|c|}{Decl(deg)}
      & \multicolumn{1}{|c|}{HPBW}
      & \multicolumn{1}{|c|}{S$_{min}$}
      & \multicolumn{1}{|c|}{N of}
      & \multicolumn{1}{|c|}{n}\\
	\multicolumn{1}{|c|}{ }
      & \multicolumn{1}{|c|}{Name}
      & \multicolumn{1}{|c|}{of}
      & \multicolumn{1}{|c|}{ }
      & \multicolumn{1}{|c|}{ }
      & \multicolumn{1}{|c|}{ }
      & \multicolumn{1}{|c|}{ }
      & \multicolumn{1}{|c|}{ }
      & \multicolumn{1}{|c|}{ }\\
	\multicolumn{1}{|c|}{(MHz)}
      & \multicolumn{1}{|c|}{ }
      & \multicolumn{1}{|c|}{publ.}
      & \multicolumn{1}{|c|}{or l(d)}
      & \multicolumn{1}{|c|}{or b(d)}
      & \multicolumn{1}{|c|}{arcmin}
      & \multicolumn{1}{|c|}{(mJy)}
      & \multicolumn{1}{|c|}{objects}
      & \multicolumn{1}{|c|}{degs$^2$}\\ \hline
}
\tabletail{\hline}
\begin{supertabular}{|rcrcccrrc|}
 10-25& UTR-2  & 78-95 &   0-24   &   $>$-13    & 25..60  & 10000 &   1754 &  0.2   \\
   38 & 8C     & 90/95 &   0-24   &   $>$+60    &   4.5   & 1000  &   5859 &  1.7   \\
   80 & CUL1   &  73   &   0-24   &    -48,+35  &   3.7   & 2000  &    999 &  0.04  \\
   80 & CUL2   &  75   &   0-24   &    -48,+35  &   3.7   & 2000  &   1748 &  0.06  \\
  151 & 6CI    &  85   &   0-24   &    $<$+80   &   4.5   &  200  &   1761 &  5.7   \\
  151 & 6CII   &  88   &  8.5-17.5&    +30,+51  &   4.5   &  200  &   8278 &  4.1   \\
  151 & 6CIII  &  90   &  5.5-18.3&    +48,+68  &   4.5   &  200  &   8749 &  4.5   \\
  151 & 6CIV   &  91   &   0-24   &    +67,+82  &   4.5   &  200  &   5421 &  3.8   \\
  151 & 6CVa   &  93   &  1.6- 6.2&    +48,+68  &   4.5   & ~300  &   2229 &  3.0   \\
  151 & 6CVb   &  93   & 17.3-20.4&    +48,+68  &   4.5   & ~300  &   1229 &  2.6   \\
  151 & 6CVI   &  93   & 22.6- 9.1&    +30,+51  &   4.5   & ~300  &   6752 &  2.7   \\
  151 & 7CI    &  90   & (10.5+41)&    (6.5+45) &   1.2   &   80  &   4723 &  9.7   \\
  151 & 7CII   &  95   &   15-19  &    +54,+76  &   1.2   & ~100  &   2702 &  6.5   \\
  151 & 7CIII  &  96   &   9-16   &    +20,+35  &   1.2   & ~150  &   5526 &  4.0   \\
  160 & CUL3   &  77   &   0-24   &    -48,+35  &   1.9   & 1200  &   2045 &  0.08  \\
  178 & 4C     &  65   &   0-24   &    -7,+80   & 15x7.5  & 2000  &   4844 &  0.2   \\
  232 & MIYUN  &  96   &   0-24   &    +30,+90  &   3.8   & ~100  &  34426 &  3.3   \\
  325 & WENSS  &  98   &   0-24   &    +30,+90  &   0.9   &   18  & 229420 &  ~22   \\
  327 & WSRT   &  91   &  5 fields&   (+40,+72) &  ~1.0   &    3  &   4157 &  ~50   \\
  365 & TXS    &  96   &   0-24   &  -35.5,+71.5&   ~.1   &  250  &  66841 &  ~2.   \\
  408 & MRC    & 81/91 &   0-24   &   -85,+18.5 &   ~3    &  700  &  12141 &  0.5   \\
  408 & B2     & 70-73 &   0-24   &    +24,+40  &  3 x10  &  250  &   9929 &  3.1   \\
  408 & B3     &  85   &   0-24   &    +37,+47  &  3 x 5  &  100  &  13354 &  5.2   \\
  408 & MC1    &  73   &   1-17   &    -22,-19  &   2.7   &  100  &   1545 &  2.3   \\
  408 & MC4    &  76   &   0-18   &    -74,-62  &   2.7   &  130  &   1257 &  1.0   \\
  608 & WSRT   &  91   &sev.fields&   (~40,~72) &   0.5   &    3  &  1693  &(~50)   \\
  611 & NAIC   &  75   &   22-13  &    -3,+19   &  12.6   &  350  &   3122 &  0.6   \\
 1400 & GB     &  72   &   7-16   &    +46,+52  &  10x11  &   90  &   1086 &  2.0   \\
 1400 & GB2    &  78   &   7-17   &    +32,+40  &  10x11  &   90  &   2022 &  2.2   \\
 1400 & WB92   &  92   &   0-24   &    -5,+82   &  10x11  & ~150  &  31524 &  0.7   \\
 1400 & NVSS39 &  98   &   0-24   &    -40,+90  &   0.9   &  2.0  &1814748 &  58.   \\
 1400 & FIRST5 &  98   &  7.3,17.4&   22.2,57.6 &   0.1   &  1.0  & 382892 &  73.   \\
 1400 & FIRST5 &  98   & 21.3,3.3 &   -11.5,+1.6&   0.1   &  1.0  &  54537 &  73.   \\
 1400 & PDF    &  98   & 1.1-1.3  &    -46,-45  & 0.1-0.2 &  0.1  &   1079 &  343   \\
 2700 & PKS    & (90)  &   0-24   &    -90,+27  &   ~8    &   50  &   8264 &  0.3   \\
 3900 & Z      &  89   &   0-24   &     0,+14   & 1.2x52  &   50  &   8503 &  1.7   \\
 3900 & RC     & 91/93 &   0-24   &    4.5,5.5  & 1.2x52  &    4  &   1189 &  3.2   \\
 3900 & Z2     &  95   &   0-24   &     0,+14   & 1.2x52  &   40  &   2944 &  0.6   \\
 4775 & MIT-GB &  83   &  22.3-13 &    -3,+19   &   2.8   &  ~40  &   2661 &  0.5   \\
 4850 & MG1-4  & 86-91 &   var.   &     0,+39   &  ~3.5   &   50  &  24180 & ~1.2   \\
 4850 & 87GB   &  91   &   0-24   &     0,+75   &  ~3.5   &   25  &  54579 &  2.7   \\
 4850 & BWE    &  91   &   0-24   &     0,+75   &  ~3.5   &   25  &  53522 &  2.7   \\
 4850 & GB6    &  96   &   0-24   &     0,+75   &  ~3.5   &   18  &  75162 &  3.7   \\
 4850 & PMNM   &  94   &   0-24   &    -88,-37  &   4.9   &   25  &  15045 &  1.8   \\
 4850 & PMN-S  &  94   &   0-24   &   -87.5,-37 &   4.2   &   20  &  23277 &  2.8   \\
 4850 & PMN-T  &  94   &   0-24   &    -29,-9.5 &   4.2   &   42  &  13363 &  2.0   \\
 4850 & PMN-E  &  95   &   0-24   &    -9.5,+10 &   4.2   &   42  &  11774 &  1.7   \\
 4850 & PMN-Z  &  96   &   0-24   &    -37,-29  &   4.2   &   70  &   2400 &  1.0   \\
\hline
\hline
     &        &       &   $l$      &  $|b|$    &         &       &        &       \\
\hline
  31 & NEK    &  88   & 350--250 & $ <2.5$&  13x 11 & 4000  &    703 &  0.7   \\
 151 & 7C(G)  &  98   & 80--180  & $ <5.5$&   1.2   & ~100  &   6262 &  4.8   \\
 327 & WSRTGP &  96   &  43--91  & $ <1.6$&  ~1.0   &  ~10  &   3984 &  ~25   \\
1400 & GPSR   &  90   &  20--120 & $ <0.8$&  0.08   &   25  &   1992 &  8.9   \\
1408 & RRF    &  90   & 357--95.5 &$ <4.0$&   9.4   &   98  &    884 &  1.1   \\
1420 & RRF    &  98   & 95.5--240 &$-4 - 5$&   9.4   &   80  &   1830 &  1.5   \\
1400 & GPSR   &  92   & 350--40  & $ <1.8$&  0.08   &   25  &   1457 &  8.1   \\
2700 & F3R    &  90   & 357--240 & $ <5  $&   4.3   &   40  &   6483 &  2.7   \\
4875 & ADP79  &  79   & 357--60  & $ <1 $&   2.6   & ~120  &   1186 &  9.4   \\
5000 & GT     &  86   & 40--220  & $ <2 $&   2.8   &   70  &   1274 &  1.8   \\
5000 & GPSR   &  94   & 350--40  & $ <0.4$&  ~0.07  &    3  &   1272 &  26.   \\
5000 & GPSR   &  79   & 190--40  & $ <2 $&   4.1   &  260  &    915 &  1.1   \\
\hline
\end{supertabular}
\end{center}
}

\newpage
\bigskip

\noindent
{\bf  Examples of the largest non-radio catalogues in CATS}
\noindent
\begin{table}[h]
\begin{tabular}{|cccccrr|}
\hline
$\lambda$& Name& Publ &    RA    &      Dec    &     N   & N/deg$^2$ \\
\hline
 opt  & PGC   &  89   &   0-24   &    -90,+90  &   73197 &  3.6   \\
 opt  & MCG   &  75   &   0-24   &    -33.5,+90&   31886 &        \\
 ir   &IRASPSC&  87   &   0-24   &    -90,+90  &  245889 & 11.9   \\
 ir   &IRASFSC&  89   &   0-24   & $|b|>10$    &  235935 & 13.8   \\
 ir   &IRASSSC&  89   &   0-24   &    -90,+90  &   43886 &        \\
 Xray & ROSAT &  95   &   0-24   &    -90,+90  &   74301 &  3.6   \\
\hline
\end{tabular}
\end{table}

We are grateful to Vladimir Chernenkov for his help with the software
design and for organizing the remote access systems for CATS.

\end{document}